\documentclass[twocolumn,notitlepage,prl,superscriptaddress,showpacs]{revtex4-1}
\usepackage{hyperref}
\usepackage[usenames,dvipsnames]{color}
\usepackage{amsmath}
\usepackage[english]{babel}
\usepackage{amssymb}
\usepackage{graphicx}
\usepackage{epstopdf}
\usepackage{tabularx}

\begin{document}


\title{Determinant Monte Carlo for irreducible Feynman diagrams in the strongly correlated regime}

\author{Fedor \v{S}imkovic IV.} \altaffiliation[Present address: ]{Centre de Physique Th\'eorique, \'Ecole Polytechnique,
CNRS, Universit\'e Paris-Saclay, 91128 Palaiseau, France and Coll\`{e}ge de France, 11 place Marcelin Berthelot, 75005 Paris, France.}
\affiliation{Department of Physics, King's College London, Strand, London WC2R 2LS, UK}

\author{Evgeny Kozik}
\affiliation{Department of Physics, King's College London, Strand, London WC2R 2LS, UK}

\date{\today}

\begin{abstract}
We develop a numerically exact method for the summation of irreducible Feynman diagrams for fermionic self-energy in the thermodynamic limit. The technique, based on the Diagrammatic Determinant Monte Carlo and its recent extension to connected diagrams, allows us to reach high ($\sim 10$) orders of the weak-coupling expansion for the self-energy of the two-dimensional Hubbard model. Access to high orders reveals a non-trivial analytic structure of the self-energy and enables its controlled reconstruction with arbitrary momentum resolution in the nonperturbative regime of essentially strong correlations, which has recently been reached with ultracold atoms in optical lattices. 
\end{abstract}


\maketitle

Definitive answers to key questions about various forms of collective behavior of interacting electrons---from quantum magnetism to photovoltaics and high-temperature superconductivity---hinge upon our ability to describe their properties reliably, i.e. without having to introduce uncontrolled systematic errors. This understanding has lead to a surge of interest in development of unbiased computational approaches for correlated fermions and the problem of controlling their error bars, which has become the focus of consorted effort (see, e.g., Ref.~\cite{LeBlanc2015_PRX} and references therein). Such systematic studies reinforce the view that there is no single universal technique that could access all aspects of correlation physics in different regimes at once. 

Quantum Monte Carlo techniques on a lattice \cite{DQMC1981, rubtsov2003, Rubtsov2005determinants, Burovski2006_NJP} are very powerful at fermion densities around one particle per site (half filling), and moderate coupling, but struggle to control the error bars in the thermodynamic limit at low temperatures and nonzero doping. Controllable approaches based on systematic extensions of the dynamical mean-field theory \cite{georges1996dmft, CDMFT2001, maier2005dca} are particularly effective whenever the observable is not sensitive to long-range correlations. Diagrammatic Monte Carlo (DiagMC) techniques \cite{van2010diagrammatic, kozik2010diagrammatic, VanHoucke2012EOS, deng2015emergent, Rossi2016det} stochastically sum all Feynman diagrams to a high order immediately in the thermodynamic limit and can reliably capture non-trivial spatial correlations, but their controllability depends on convergence properties of the series, which typically diverges already at moderate interaction strengths. The current state of the field is that perhaps the most interesting regime of moderate-to-strong interactions, which is expected to harbour non-trivial correlation physics at relatively high temperatures, is hardly under control by any available computational method.

With the lack of accurate theoretical solutions, a promising approach is experimental emulation of basic models of correlated electrons in solids, in particular with ultracold atoms loaded in an optical lattice~\cite{jaksch1998cold, Bloch_review_2005, kohl2005fermionic, lewenstein2007ultracold,jordens2008mott, schneider2008metallic,hulet2015antiferromagnetism, greif2015formation,parsons2016site,greiner2017}. This field has seen dramatic progress since the realisation~\cite{kohl2005fermionic,jordens2008mott, schneider2008metallic} of the prototypical fermionic Hubbard model~\cite{hubbard1963electron,anderson1963theory,anderson1997theory,Lee2006}
\begin{align}
	H = -  t\sum_{\left<i,j\right> \sigma}  \hat{c}^{\dagger}_{i, \sigma} \hat{c}_{j, \sigma}
+ U \sum_{i} \hat{n}_{i, \uparrow} \hat{n}_{i \downarrow} -\sum_{i, \sigma} \mu_\sigma \hat{n}_{i, \sigma},
    \label{Hubbard}
\end{align}
where $\mu_\sigma$ is the chemical potential, $\hat{c}^{\dagger}_{i, \sigma}$ and $\hat{c}_{i, \sigma}$
create and annihilate (respectively) a fermion with the spin $\sigma \in \{\uparrow, \downarrow\}$
on the site $i$, and $\hat{n}_{i, \sigma}=\hat{c}^{\dagger}_{i, \sigma} \hat{c}_{i, \sigma}$. Until recently, this model has been studied at relatively high temperatures, where several theoretical approaches provide reliable bechmarks for calibration and cross-validation of results. However, substantial progress of cooling and probing techniques has already allowed access to sufficiently low temperatures to observe magnetic properties~\cite{hulet2015antiferromagnetism, greif2015formation,parsons2016site}, and, very recently, detect long-range antiferromagnetic correlations in the 2D Hubbard model at temperatures as low as $T=0.25t$ and $U\approx 7t$~\cite{greiner2017}. The experiments have thus already reached the most challenging regime for all current theoretical methods. 

In this Rapid Communication, we introduce a numerically exact approach for the stochastic summation of irreducible Feynman diagrams for the fermionic self-energy based on diagrammatic determinant Monte Carlo (DDMC)~\cite{rubtsov2003, Rubtsov2005determinants, Burovski2006_NJP} and its recent extension to connected diagrams in the thermodynamic limit (CDet)~\cite{Rossi2016det}, $\Sigma$DDMC. The method allows us to reach high orders of the diagrammatic series for the self-energy of the two-dimensional Hubbard model (\ref{Hubbard}) immediately in the thermodynamic limit. Although the series manifestly diverges in strongly correlated regimes, access to high orders enables a systematic protocol for reconstructing the self-energy with controlled accuracy and arbitrary momentum resolution. We demonstrate the technique for typical parameters $U=7t$, $T=0.2t$, $\mu=2t$ (density $n  = 0.950(6)$, we set $t=1$ below), the regime where other methods are currently struggling to reach a controlled solution~\cite{LeBlanc2015_PRX}, and use it to obtain the corresponding (quasi)momentum distribution, which can be observed experimentally. 

We start from the effective action with the arbitrary free parameter $\alpha$ \cite{rubtsov2003, Rubtsov2005determinants,profumo2015, wu2017} and the expansion variable $\xi$, 
\begin{eqnarray}
  &S_{\xi} =  -\sum_{\omega_{n},\mathbf{k},\sigma}    c_{\omega_{n},\mathbf{k},\sigma}^{\dagger}G^{(0)}_{\sigma}(i\omega_{n},\mathbf{k})^{-1}c_{\omega_{n},\mathbf{k},\sigma} \\
\label{eq:action}
 & - \xi \sum_{\omega_{n},\mathbf{k},\sigma} \alpha_\sigma  c_{\omega_{n},\mathbf{k},\sigma}^{\dagger} c_{\omega_{n},\mathbf{k},\sigma}
  + \xi U\int_{0}^{\beta}n_{\tau\uparrow}n_{\tau\downarrow}d\tau,
\nonumber
\end{eqnarray}
where $G^{(0)}_{\sigma}(i\omega_{n},\mathbf{k})^{-1} = i\omega_n + \mu_\sigma - \epsilon_{\mathbf{k} \sigma} - \alpha_\sigma$ and $\epsilon_{\mathbf{k} \sigma}$ is the dispersion relation of the noninteracting system. At $\xi=1$ the action corresponds to the Hamiltonian~(\ref{Hubbard}). Expansion in powers of $\xi$ and application of Wick's theorem leads to the diagrammatic series for the partition function $Z$ (see, e.g., Ref.~\cite{AGD}). The key observation \cite{rubtsov2003, Rubtsov2005determinants} is that the sum of all $(n!)^2$ diagrams of order $n$ for a given configuration $\mathcal{S}_n = \{\mathrm{x}_j , j=1,\dots,n\}$ of $n$ vertices in space-imaginary time, $\mathrm{x}_j=(\mathbf{x}_j, \tau_j)$, can be recovered from the product of two determinants 
\begin{equation}
z(\mathcal{S}_n) \!= \!\det\mathrm{A}^{\uparrow}(\mathcal{S}_n) \det\mathrm{A}^{\downarrow}(\mathcal{S}_n), \label{z_n}
\end{equation}
where the matrices $\mathrm{A}^\sigma$ are constructed from the Green's functions, $A^\sigma_{ij}(\mathcal{S}_n) = G_\sigma^{(0)}(\mathrm{x}_j-\mathrm{x}_i) - (\alpha_\sigma/U) \delta_{ij}$.
Thus, the net contribution of the factorial number of diagrams of order $n$ can be computed in only $\mathcal{O}(n^3)$ elementary operations, and the sum over all possible $\mathcal{S}_n$ can be efficiently sampled by Monte Carlo~\cite{rubtsov2003, Rubtsov2005determinants, Burovski2006_NJP}. Nonetheless, although the series for $Z$ converges for any finite volume $V$ and inverse temperature $\beta=1/T$, the average diagram order grows as $\beta V U$, and in typical interesting cases the extrapolation to the thermodynamic limit is practically impossible. This is because the sum (\ref{z_n}) is dominated by disconnected diagrams---those containing pieces that are not linked to the rest of the diagram by a $G^{(0)}$ line---responsible for the exponential scaling of $Z$ on $\beta V$.    
 
A family of DiagMC techniques, stochastically summing only the connected (and typically also irreducible~\footnote{Here, by irreducible we mean those diagrams that can not be split into two disconnected pieces by cutting any fermionic line.}) diagrams~\cite{van2010diagrammatic, kozik2010diagrammatic, deng2015emergent}, enables calculation of dynamical observables, such as the fermionic self-energy $\Sigma_{\mathrm{k} \sigma}$, $\mathrm{k}=(\mathbf{k}, - i\omega_{n})$, immediately in the thermodynamic limit. However, until recently, these diagrams have been sampled in DiagMC one by one, constituting a factorial scaling of computational time with the diagram order $n$. While diagram orders as high as $n \sim 6$ could be accessed in practice, it is typically insufficient for obtaining controlled results at strong correlations. It is therefore tempting to take advantage of the determinantal summation in the spirit of Eq.~(\ref{z_n}) for series of irreducible diagrams as well. 

It was shown recently that determinantal summation could be applied to series of \textit{connected} diagrams immediately in the thermodynamic limit~\cite{Rossi2016det}. The idea is that for each configuration of vertices $\mathcal{S}_n$ in Eq. (\ref{z_n}) one can recursively subtract all the disconnected diagrams, constructed from the determinants of the principal submatrices of $\mathrm{A}^\sigma$. While all possible diagrams of order $n$ take $\mathcal{O}(n^3)$ elementary operations to sum by a single determinant, extracting only the connected ones requires computing determinants for all proper subsets of $\mathcal{S}_n$, which is $\mathcal{O}(n^3 2^n)$ steps, and a number of subtractions that grows exponentially as $\mathcal{O}(3^n)$~\cite{Rossi2016det}. Nonetheless, it beats the factorial scaling of DiagMC in the theoretical large-$n$ limit. Most importantly, it was shown in Ref.~\cite{Rossi2016det} that in practice this trick allows to reach diagram orders as high as $n\sim 10$ for the grand potential density and obtain the pressure with unmatched accuracy in the weakly correlated regime ($U \sim 2$). The approach of determinantal summation/subtraction can be extended to sums of irreducible diagrams for the self-energy, as we shall see, at an exponential cost as well. The question remains, however, of whether it can bring any practical benefits in terms of enabling access to the regime of truly strong correlations.   

The series for the self-energy $\Sigma_{\mathrm{k} \sigma}$ consists of all possible connected irreducible diagrams with two vertices that lack their respective incoming/outgoing propagators with spin $\sigma$ and four-(quasi)-momentum $\mathrm{k}$~\cite{AGD}. For a particular vertex configuration $\mathcal{S}_n$, we can compute the sum of all $(n!)^2$ diagrams (including disconnected and reducible) with two open ends carrying $\sigma$ and $\mathrm{k}$ by
\begin{equation}
z_{\mathrm{k}\sigma}(\mathcal{S}_n, l\!\!:\!\mathrm{x}_l)=\det\mathrm{A}^{\sigma}_{\mathrm{k}}(\mathcal{S}_n, l\!\!:\!\mathrm{x}_l) \det\mathrm{A}^{\bar{\sigma}}(\mathcal{S}_n), \label{z_n_k}
\end{equation}
where the matrix $\mathrm{A}^{\sigma}_{\mathrm{k}}(\mathcal{S}_n, j\!\!\!:\!\!\!\mathrm{x}) $ is obtained from $\mathrm{A}^{\sigma}(\mathcal{S}_n)$ by replacing its $j$-th column by the vector $\left(e^{\imath \mathrm{k} \left( \mathrm{x} - \mathrm{x}_1 \right)} , \; e^{\imath \mathrm{k} \left( \mathrm{x} - \mathrm{x}_2 \right)} , \; \hdots , \; e^{\imath \mathrm{k} \left( \mathrm{x} - \mathrm{x}_n \right)}\right)^{\intercal}$, and $\sigma \neq \bar{\sigma}$. Now the task is to remove from $z_{\mathrm{k}\sigma}$ all the disconnected and reducible diagrams. To this end, we define an auxiliary quantity
\begin{equation}
f_\sigma(\mathcal{S}_n, l\!\!:\!\mathrm{x})=\det\mathrm{A}^{\sigma}_{G}(\mathcal{S}_n, l\!\!:\!\mathrm{x}) \det\mathrm{A}^{\bar{\sigma}}(\mathcal{S}_n), \label{f}
\end{equation}
with the matrix $\mathrm{A}^{\sigma}_G(\mathcal{S}_n, l\!\!\!:\!\!\!\mathrm{x})$ obtained from $\mathrm{A}^{\sigma}(\mathcal{S}_n)$ by replacing $\mathrm{x}_l$ in its $l$-th column by $\mathrm{x}$. In essence, $f_\sigma$ sums all the diagrams of the general structure 
$z \Sigma(\mathrm{x}_{i_1}-\mathrm{x}_l) G^{(0)}(\mathrm{x}_{i_2}-\mathrm{x}_{i_1}) \ldots \Sigma(\mathrm{x}_{i_{p}}-\mathrm{x}_{i_{p-1}}) G^{(0)}(\mathrm{x}-\mathrm{x}_{i_p})$, which start at the self-energy vertex $\mathrm{x}_l$ from the set $\mathcal{S}_n$, end with a propagator going to the external vertex $\mathrm{x}$, and may have disconnected parts $z$. 
Thus, the sum of all self-energy diagrams for the configuration $\mathcal{S}_n$ can be obtained recursively from the formula:
\begin{eqnarray}
s_{\mathrm{k}\sigma} (\mathcal{S}_n,  l\!\!:\!\mathrm{x}_l)=z_{\mathrm{k}\sigma}(\mathcal{S}_n, l\!\!:\!\mathrm{x}_l) -  
\!\!\!\!\!\!\!\!\!\! \sum_{\mathcal{S}_p \subset \mathcal{S}_n: \mathrm{x}_l \in \mathcal{S}_p}  \!\!\!\!\!\!\!\!\!\!  z(\mathcal{S}_n \!\! \setminus \! \mathcal{S}_p) s_{\mathrm{k}\sigma}(\mathcal{S}_p, l\!\!:\!\mathrm{x}_l)  \nonumber \\
- \!\!\!\!\!\!\! \sum_{\mathcal{S}_p \subset \mathcal{S}_n: \mathrm{x}_l\in \mathcal{S}_p} \; \sum_{ \mathrm{x}_m \in (\mathcal{S}_n \setminus \mathcal{S}_p)} \!\!\! f_{\sigma}(\mathcal{S}_p, l\!\!:\!\mathrm{x}_m) \; s_{\mathrm{k}\sigma}(\mathcal{S}_n \!\! \setminus \!\mathcal{S}_p, m\!\!:\!\mathrm{x}_l) \;\;\;\;\;\;   \label{s}
\end{eqnarray}  
Here $\mathrm{x}_l \in \mathcal{S}_n$, the first sum is over proper subsets $\mathcal{S}_p$ of $\mathcal{S}_n$ that include $\mathrm{x}_l$, the last term is additionally summed over all vertices $\mathrm{x}_m$ that belong to $\mathcal{S}_n$ but not to $\mathcal{S}_p$. 

Finally, the expansion of $\Sigma_{\mathrm{k} \sigma}$ in powers of $\xi$ (we restore the explicit dependence on external parameters) reads
\begin{equation}
\Sigma_{\mathrm{k} \sigma}(T, \mu, U) = \sum_{n=1}^{\infty} a_{n, \mathrm{k}\sigma}(T, \mu_\sigma-\alpha_\sigma, \alpha_\sigma/U) (U \xi)^n , \label{Sigma_series}
\end{equation}
with the coefficients
\begin{equation} 
a_{n, \mathrm{k}\sigma}=\frac{(-1)^n}{n!} \sum_{\mathcal{S}_n, l} s_{\mathrm{k}\sigma} (\mathcal{S}_n, l\!\!:\!\mathrm{x}_l), \label{a_n}
\end{equation}
where $\sum_{\mathcal{S}_n} = \sum_{\mathbf{x}_1 \dots \mathbf{x}_n} \prod_{j=1}^{n} \int_{0}^{1/T} d\tau_j$. 
The sum over all vertex configurations can be efficeintly computed by the standard continous-time Metropolis-type scheme (see, e.g. Ref.~\cite{DDMC_Neel} for details). At each Monte Carlo step, the evaluation of $s_{\mathrm{k}\sigma}$ is done in two stages: First, all determinants involved in Eq.~(\ref{s}) are computed in $\mathcal{O}( n^5 2^n)$ elementary operations, then the recursive procedure (\ref{s}) is performed in $\mathcal{O}(n^2 3^n)$ steps. This scaling could be improved: the tree algorithm \cite{griffin2006minors, simkovic2018thesis} and fast subset convolution~\cite{bjorklund2007subset_convolution} reduce these costs to $\mathcal{O}( n^2 2^n)$ and $\mathcal{O}(n^4 2^n)$ [$\mathcal{O}(2^n)$ and $\mathcal{O}(n^2 2^n)$ for CDet~\cite{simkovic2018thesis}], respectively. For realistically accessible orders ($n \sim 12$) the computational cost of both approaches is comparable. Given $\mathcal{S}_n$, our code evaluates the sum of all $24 \, 936 \, 416$ irreducible diagrams at $n=10$~\cite{simkovic2018thesis} averaged over $10!$ permutations of vertices in $\sim 0.01s$ on modern CPU.

\begin{figure}[htbp]
\vspace*{0.0cm}
\includegraphics[width=1.\columnwidth]{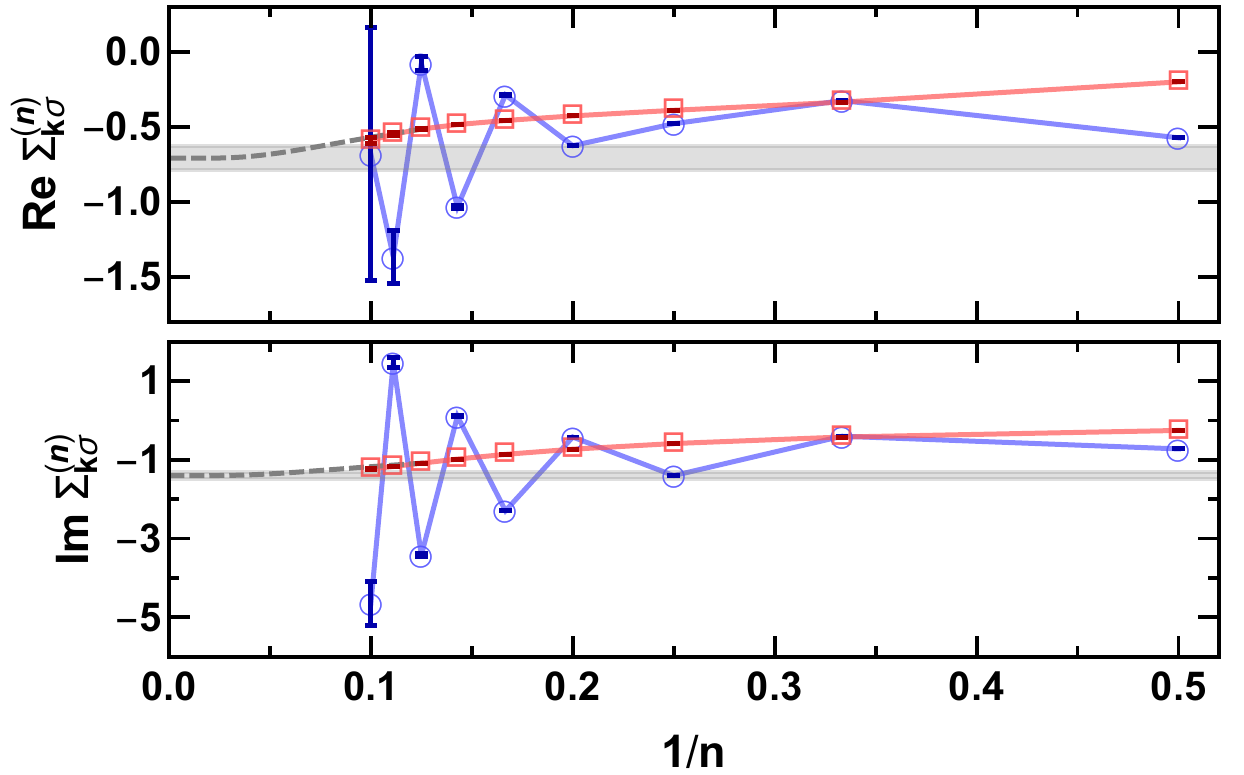}
\vspace*{0.0cm}
\caption{\label{fig_Sigma} (color online) Partial sum $\Sigma_{\mathrm{k} \sigma}^{(n)}$ of the original expansion~(\ref{Sigma_series}) at $\xi=1$ (circles) and that in terms of the transformed variable $w(\xi)$ (squares) at $T=0.2t$, $U=7t$, $\mu=2t$, $\omega=\omega_0$, $\mathbf{k}=(\pi/8, \pi)$ as a function of inverse truncation order. The dashed line is an extrapolation by the series for $g(w(1))$, Eq.~(\ref{IA}). The horizontal bands are the claimed result, Fig.~\ref{fig_IA}. 
}
\end{figure}

\begin{figure}[htbp]
\vspace*{0.0cm}
\includegraphics[width=1.\columnwidth]{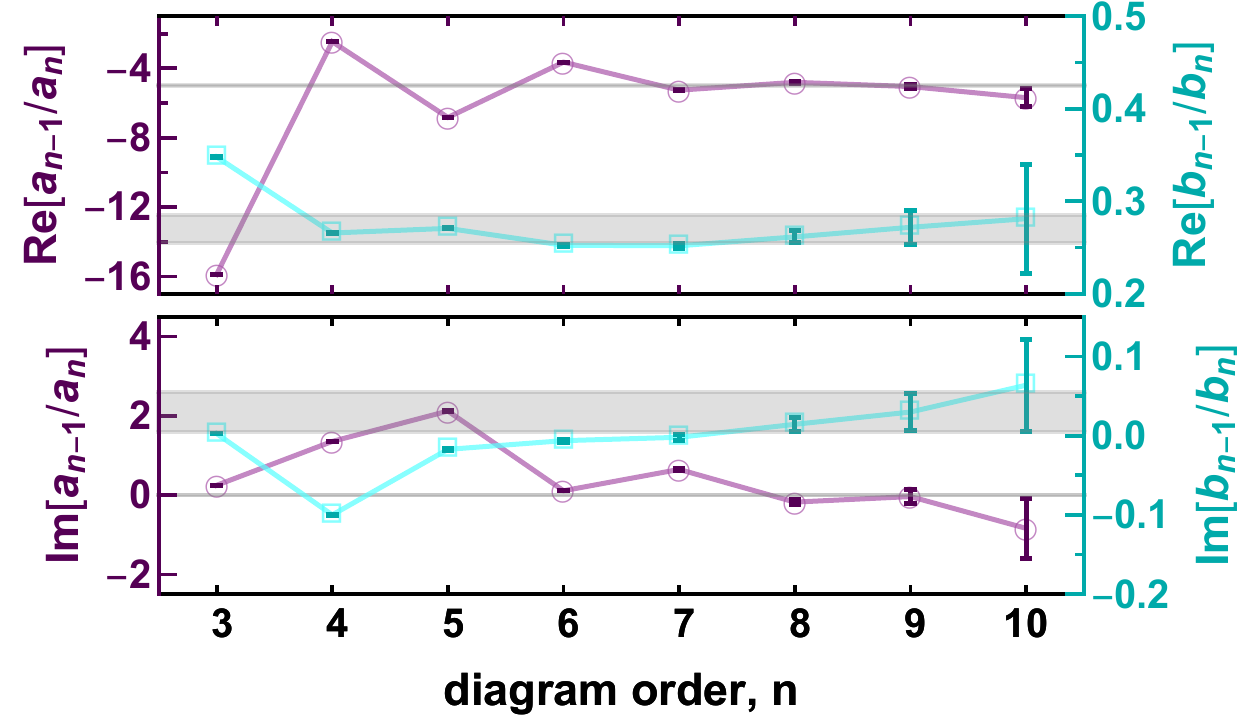}
\vspace*{0.0cm}
\caption{\label{fig_an_bn} (color online) The ratio of successive coefficients of the original, $a_{n-1}/a_n$ (circles), and transformed by the map $w(\xi)$, $b_{n-1}/b_n$ (squares), series for the parameters of Fig.~\ref{fig_Sigma}. The first singularity is observed at $ U\xi_{s_1} \approx -5$ and the second one at $w_{s_2} \approx 0.27+i 0.03$ (corresponding to $ U\xi_{s_2} \approx -9$), while $w(\xi=1)=0.23$.}
\end{figure}

We now turn to the problem of reconstructing $\Sigma_{\mathrm{k} \sigma}$ given the series coefficients $a_n$ obtained by $\Sigma$DDMC. 
Note that, by construction, $\Sigma_{\mathrm{k} \sigma}$ does not depend on the choice of $\alpha$, but the series is different for each $\alpha$, which can be used, e.g., to control its convergence~\cite{profumo2015, wu2017}. Here we use this freedom to maximise the order $n$ we can reach, which empirically amounts to nullifying the diagonal of $\mathrm{A}^{\sigma}$, so that $\alpha$ is found from $G_\sigma^{(0)}(\mathbf{x}=0, \tau=-0) = \alpha/U$. For method validation, we have reproduced current state-of-the-art benchmarks~\cite{LeBlanc2015_PRX, wu2017}. Here, we address an essentially correlated regime, where controlled results for $\Sigma_{\mathrm{k} \sigma}$ in the thermodynamic limit are not accesible by other methods~\cite{LeBlanc2015_PRX}. The result of the partial sum $\Sigma^{(n)}=\sum_{m=0}^n a_m U^m$ at the lowest Matsubara frequency $\omega_0$ and $\mathbf{k}=(\pi/8, \pi)$ as a function of $n$ up to ${n_{\mathrm{max}}}=10$, shown in Fig.~\ref{fig_Sigma}, evidences that the series is wildly divergent. It is known, however, \cite{Benfatto2006, van2010diagrammatic, kozik2010diagrammatic} that the series (\ref{Sigma_series}) at $T>0$ generally has a non-zero convergence radius. Except for special cases, the position of the singularity closest to the origin in the complex plane of the expansion parameter $\xi$ can be found from the ratio test, $ U \xi_{s1} = \lim_{n \to \infty} a_{n-1}/a_n$. Fig.~\ref{fig_an_bn} shows $a_{n-1}/a_n$ as a function of $n$, suggesting $ U \xi_{s1} \approx -5$. 
%
\begin{figure}[htbp]
\vspace*{0.0cm}
\includegraphics[width=1.\columnwidth]{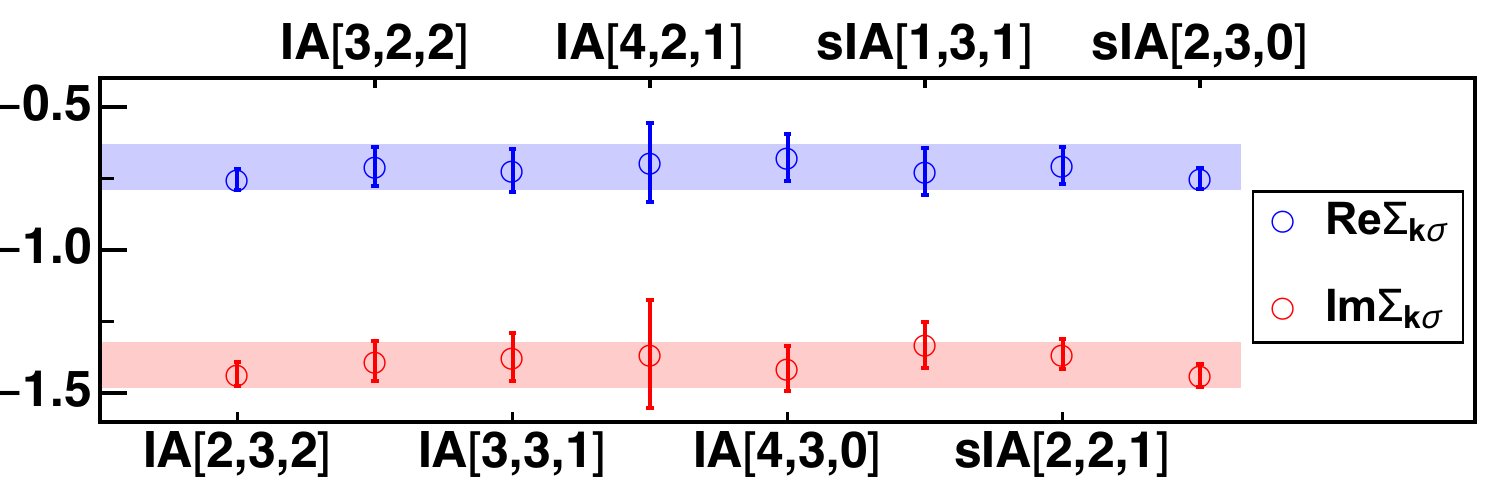}
\vspace*{0.0cm}
\caption{\label{fig_IA} (color online) Evaluation of $\Sigma_{\mathrm{k} \sigma}$ for the parameters of Fig.~\ref{fig_Sigma} using the IA method~(\ref{IA}) applied to the series~(\ref{Sigma_series}) (labelled IA) and the shifted series for $\Sigma/\xi^2$ (sIA) for various choices of $[L,M,N]$. The horizontal bands show the claimed result with the error bar.}
\end{figure}

The singularity with $\mathrm{Re}\xi_{s_1}<0$ is an inconvenience, but does not prevent one from accurately obtaining the self-energy at $\xi=1$. A standard approach is analytic continuation based on conformal maps. Given $\Sigma(\xi)$, which is analytic at $\xi=0$ in the open disk $|\xi|<|\xi_{s_1}|$, the idea is to transform the complex plane of $\xi$ using an analytic function $w=w(\xi)$, $w(0)=0$, to a domain of the complex variable $w$ where the singularity is farther away from the origin than the image of $\xi=1$, $|w(\xi_{s_1})|>|w(1)|$. As a function of $w$, $\Sigma(\xi(w))$ is then analytic at $w=0$ in the open disk $|w|<|w(\xi_{s_1})|$, which now contains $w(1)$. Re-expanding $\sum_n a_n (\xi(w))^n$ in powers of $w$, we obtain $\Sigma(\xi)= \sum_n b_n (w(\xi))^n$, which converges at $\xi=1$. 

Such a map is not unique. We choose $\xi=-4\xi_{s_1} w/(1 -w)^2$, which maps the disk $|w|<1$ onto the complex plane of $\xi$ with a branch cut along the real axis from $\xi_{s_1}$ to $-\infty$. The coefficients $b_n$ determine the position of the next singularity nearest to the origin $w_{s_2}$. 
The plateau of $b_{n-1}/b_n$ at $n \gtrsim 7$ (Fig.~\ref{fig_an_bn}) gives $w_{s_2}$ and confirms that the expansion in $w$ is indeed convergent,  $|w_{s_2}| >|w(1)|$. This observation is key for a controlled extrapolation of $\Sigma$ w.r.t. $n \to \infty$, which would be impossible with current DiagMC, typically cut off at $n\sim 6$. Depending on the map, singularities other than $\xi_{s_2}$ can appear closer to the origin and will manifest themselves in $b_{n-1}/b_n$. The configuration of singularities generally changes with $\mathrm{k}$.

To evaluate the series, we use the integral approximant~\footnote{Sometimes called Differential Approximant.} (IA) technique~\cite{hunter1979IA}, which associates the result with the function $g(w)$, $\Sigma=g(w(1))$, that has the same Taylor series as $\Sigma(w)$ up to the highest accessible order ${n_{\mathrm{max}}}$ and satisfies the differential equation
\begin{equation}
Q_M(x) g'(x) + P_L(x) g(x) + R_N(x) = 0. \label{IA}
\end{equation}
Here $Q, P, R$ are polynomials of orders $M, L, N$, respectively, determined as the unique solution of Eq.~(\ref{IA}) for $g(x)=\sum_{n=0}^{n_{\mathrm{max}}} b_n x^n$ up to terms $\mathcal{O} (x^{M+N+L+2})$ with $M+N+L+2=n_{\mathrm{max}}$. In effect, Eq.~(\ref{IA}) continues the (not necessarily convergent) series for $g(x)$ from $n \leq n_{\mathrm{max}}$ to infinite order and reconstructs the function behind it. The IA approach reduces to other standard resummation methods, such as, e.g., Pad\'{e}~\cite{Brezinski1996Pade} and Dlog-Pad\'{e}~\cite{baker1961Dlog}, as special cases~\cite{hunter1979IA}, capturing a more general analytic structure with algebraic singularities: Near the singular point $x_s$ [a zero of $Q_M(x)$] $g(x)=\phi_1(x) (1-x/x_s)^{-\nu} +  \phi_2(x)$, with functions $\phi_{1,2}(x)$ regular at $x_s$. 

The resummation (\ref{IA}) can also be used to obtain $\Sigma$ or $\Sigma/\xi^2$ (since $a_0=a_1=0$) directly from the divergent series (\ref{Sigma_series}) (Fig.~\ref{fig_IA}). In this case, $Q, P, R$ are constructed for $g(x)=\sum_{n=0}^{n_{\mathrm{max}}} a_n x^n$ or $g(x)=\sum_{n=2}^{n_{\mathrm{max}}} a_n x^{n-2}$, respectively. We verify that the bias introduced by the extrapolation (\ref{IA}) is negligible by observing that the discrepancy between the estimates of $\Sigma$ obtained for different appropriate~\cite{hunter1979IA} choices of $[L,M,N]$ is negligible compared to the corresponding statistical error. Vice versa, a measurable deviation form the assumed asymptotic form (\ref{IA}) would manifest itself as an inconsistency between different IAs beyond error bars. The results are in perfect agreement with those obtained for the transformed series (not shown), providing further evidence that the systematic error of the adopted resummation procedure is negligible. As a by-product, the procedure yields an estimate of the nearest singularity location, $ U \xi_{s1} = [-5.7(3)-i 0.2(8)]$, which may correspond to the s-wave superfluid transition~\cite{Rossi2016det} [at a different density in view of Eq.~\ref{Sigma_series}]. Since our $\{a_n\}$ ($\{b_n\}$) have error bars, we found that approximants with more general asymptotics, such as, e.g., hypergeometric/Meijer-$G$~\cite{mera2015hypergeom, mera2018maijer_g} or Borel-Pad\'{e}/Borel-Dlog-Pad\'{e}~\cite{borel1928lecons, janke1998resummation}, result in a large uncertainty of the extrapolation unless additional constraints on the number or form of singularities are introduced~\footnote{The singularity structure of $\Sigma$ can be rather involved~\cite{wu2017} and is generally unknown}.

\begin{figure}[htbp]
\vspace*{0.0cm}
\includegraphics[width=1.0\columnwidth]{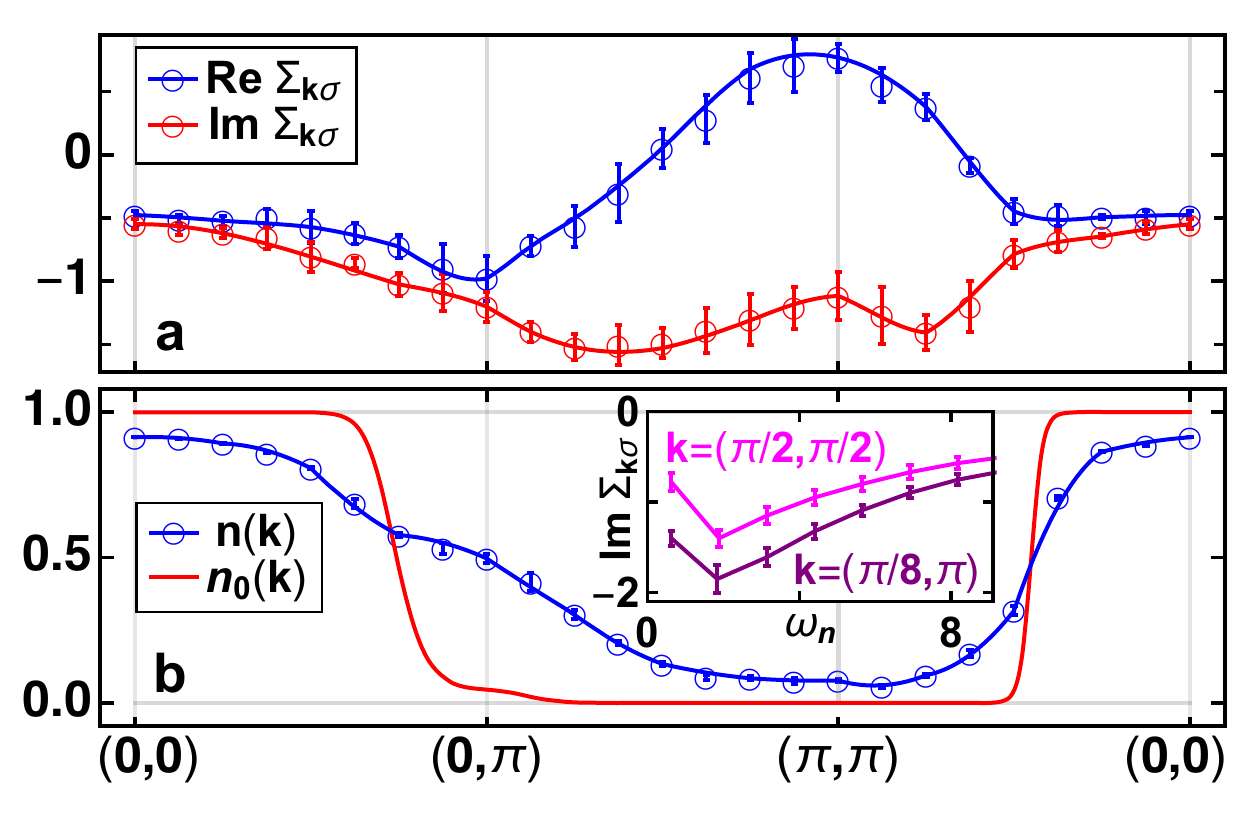}
\vspace*{0.0cm}
\caption{\label{fig_GXMG} (color online) (a) $\Sigma_{\mathbf{k} \sigma}$ at $\omega=\omega_0$; Inset: $\Sigma_{\mathbf{k} \sigma}(i\omega_n)$ at $\mathbf{k}=(\pi/2, \pi/2), (\pi/8, \pi)$ (b) the corresponding momentum distribution $n(\mathbf{k})$ at $T=0.2t$, $U=7t$, $\mu=2t$ [$n=0.950(6)$] and $n_0(\mathbf{k})$ of the ideal Fermi gas, $U=0$. }
\end{figure}

We follow this protocol to map out the momentum dependence of $\Sigma_{\mathrm{k} \sigma}$ at $\omega=\omega_0$ (Fig.~\ref{fig_GXMG}a). The knowledge of $\Sigma_{\mathrm{k} \sigma}(i\omega_n)$ allows to obtain an accurate estimate of the momentum distribution $n(\mathbf{k})=\langle c^{\dagger}_{\mathbf{k}} c_{\mathbf{k}} \rangle$ via the Dyson equation~\cite{AGD} (Fig.~\ref{fig_GXMG}b). We found that this approach leads to a more accurate estimate for $n(\mathbf{k})$ than computing it directly with CDet. The shape of $n(\mathbf{k})$ is qualitatively different from that of the corresponding non-interacting Fermi gas ($U=0$), revealing that the system is of strongly-correlated non-Fermi liquid character: The shoulder around $(0, \pi)$ is due to the breakdown of the condition $\mathrm{Im} \Sigma_{\mathbf{k} \sigma}(i \omega_n) \propto \omega_n$ for small $\omega_n$, seen, e.g, for $\mathbf{k}=(\pi/8,\pi)$ in the inset of Fig.~\ref{fig_GXMG}b. The function $n(\mathbf{k})$ can be straightforwardly probed experimentally with ultracold atoms in optical lattices~\cite{Bloch_review_2005}, which have recently been brought to this regime of parameters~\cite{greiner2017}. Our data thus provide a controlled theoretical benchmark for the ongoing studies of strong correlations in the 2D Hubbard model.

In conclusion, we note that, following Ref.~\cite{rossi2017polynomial}, the exponential computational cost of $\Sigma$DDMC with the observation that the transformed series converges implies that the computational time scales polynomially with the inverse of the desired error bar, which is generally unattainable in finite-system-size methods due to the negative sign problem~\cite{troyer2005sign}. In addition to controlled determination of observables in regimes analytically connected to the non-interacting limit, the diagrammatic approach offers a unique means of detecting and analysing phase transitions. Being fundamentally free from finite-size effects, the series (\ref{Sigma_series}) is bound to diverge at a point of non-analyticity. The key result is that $n \sim 10$ accessed by $\Sigma$DDMC appears to be in the asymptotic regime at least at $T \sim 0.2$, meaning that the point of non-analyticity and potentially certain critical properties can be found from the analysis of $a_n$ suggested above. 

\textit{Note added.} An algorithm similar to Eqs.~(\ref{z_n_k})-(\ref{a_n}) was introduced recently in Ref.~\cite{moutenet2018determinant} and applied in a regime where the series converges. An alternative approach was subsequently proposed in Ref.~\cite{rossi2018sigma}. $\Sigma$DDMC was later used for a controlled description of the metal-to-insulator crossover in the half-filled 2d Hubbard model~\citep{Fedor:2018vo}. 

\begin{acknowledgments}
We are grateful to Michel Ferrero and Alice Moutenet for discussions of the algorithm and to Aaram J. Kim, H\'{e}ctor  Mera and Branislav Nikoli\'{c} for discussions of resummation techniques. This work was supported by the Simons Foundation as a part the Simons Collaboration on the Many Electron Problem and by EPSRC through Grant No. EP/P003052/1.
\end{acknowledgments}

\bibliography{refs.bib}

\end{document}